\documentclass[twocolumn]{aastex631}
\usepackage{comment}
\usepackage{xcolor}
\usepackage{float}

\begin{document}

\title{Effects of Varied Cosmic Ray Feedback from AGN on Massive Galaxy Properties}

\author[0009-0008-8670-3386]{Charvi Goyal}
\affiliation{TAPIR, California Institute of Technology, Mailcode 350-17, 
Pasadena, CA 91125, USA}

\author[0000-0002-7484-2695]{Sam B. Ponnada}
\affiliation{TAPIR, California Institute of Technology, Mailcode 350-17, 
Pasadena, CA 91125, USA}

\author[0000-0003-3729-1684]{Philip F. Hopkins}
\affiliation{TAPIR, California Institute of Technology, Mailcode 350-17, 
Pasadena, CA 91125, USA}

\author[0000-0002-3977-2724]{Sarah Wellons}
\affiliation{Department of Astronomy, Van Vleck Observatory, Wesleyan University, 96 Foss Hill Drive, Middletown, CT 06459, USA 
}

\author[0000-0003-1896-0424]{Jose A. Benavides}
\affiliation{Department of Physics and Astronomy, University of California, Riverside, 900 University Avenue, Riverside, CA 92521, USA}

\author[0000-0003-1598-0083]{Kung-Yi Su}
\affiliation{Department of Physics \& Astronomy and Center for Interdisciplinary Exploration and Research in Astrophysics(CIERA), Northwestern University, 1800 Sherman Ave, Evanston, IL 60201, USA}



\begin{abstract}

Active galactic nuclei (AGN) provide energetic feedback necessary to `turn off' star formation in high-mass galaxies (M$_{\rm halo} \geq $ 10$^{12.5}$ M$_{\odot}$, $10.4 \leq \log(\frac{M_*}{M_\odot}) \leq 11$) as observed. Cosmic rays (CRs) have been proposed as a promising channel of AGN feedback, but the nature of CR feedback from AGN remains uncertain. We analyze a set of high-resolution simulations of massive galaxies from the Feedback in Realistic Environments (FIRE-3) project including multi-channel AGN feedback, explicitly evolving kinetic/mechanical, radiative, and spectrally-resolved CRs from the central black hole. Specifically, we explore different CR feedback and transport assumptions, calibrated to Milky Way local ISM constraints, and compare them to observed galaxy scaling relations. We find that all parameterizations explored self-regulate within agreement with observed galaxy scaling relations, demonstrating that CR injection efficiencies varied by $\sim$1.5 dex and locally-variable transport produce quenched galaxies with reasonable bulk properties; however, they feature orders-of-magnitude variant circumgalactic medium (CGM) gas properties. Our results indicate that multi-wavelength synthetic observations probing these varied halo properties from larger simulated samples in conjunction with observational comparisons may place novel constraints on how AGN physically quench star formation in massive galaxies.

\end{abstract}



\section{Introduction} \label{sec:intro}
The ``quenching problem'' of massive galaxies has been a longstanding discrepancy in galaxy formation between observed star formation rates and those predicted by simulations in massive galaxies \citep{Cole_2002,Springel_2003,Longair:2008gba,Keres_2009,Benson_2010,hopkins_fire-3_2023}. Galaxies with high dark matter halo masses (M$_{\rm halo}$ $\gtrsim 10^{12.5}$  M$_{\odot}$) appear ``red and dead" and are observed to have less gas cooling than predicted \citep{Benson_2010}. It is well-established that feedback processes play a vital role in preventing gas from cooling down and collapsing into stars \citep{SilkandRees1998,Hopkins2004,Croton2006,Croton2008,Longair:2008gba,Benson_2010,Hopkins_2018}, rectifying this `overcooling problem; however, the specific mechanisms by which quenching occurs remain unclear.

Due to their immense mass and energy output, Active Galactic Nuclei (AGN) are thought to be critical suppressors of gas cooling in the interstellar medium (ISM), circum-galactic medium (CGM) and intracluster medium (ICM) \citep{SilkandRees1998,Croton2006,Longair:2008gba,  Somerville_&_Dave_2015, Su_2021}. Several studies have found that models with AGN feedback channels are necessary to explain massive galaxy properties when compared to non-AGN counterparts \citep{Harrison_2018, su_failure_2019, Su_2021, hopkins_fire-3_2023, wellons_exploring_2023}, which is why most modern simulations which aim to reproduce the observed bulk properties of galaxies in a statistical manner include AGN feedback \citep{Schaye_2014, Sijacki_2015, Somerville_&_Dave_2015, Dav_2019, su_failure_2019}.

How energy from AGN is transferred and coupled to the gas of the host galaxy is now the major open question. Motivated by observations, most of the theoretical literature has focused on the radiative and mechanical feedback modes of AGN \citep[see][for a recent review]{harrison_observational_2024}. Observed UV and X-ray emission from AGN suggest that AGN generate strong radiation fields due to accretion, which can Compton-heat, photo-ionize, and drive winds via radiation pressure. AGN also drive mechanical outflows and jets \citep[for reviews, see][]{fabian_observational_2012,heckman_coevolution_2014}, which can evacuate large cavities and thermalize energy via shocks and turbulent mixing. 

AGN are also prodigious sources of cosmic rays (CRs), seen in synchrotron, $\gamma$-ray, and inverse Compton emission, which could affect surrounding gas via ionization, heating, or large-scale non-thermal pressure gradients in galactic halos  \citep{butsky_role_2018,Su_2019,Hopkins2020, bustard_cosmic-ray_2021}. 
How these different channels influence quenching is a topic of ongoing study in galaxy formation \citep{Benson_2010, Somerville_&_Dave_2015, Su_2019,Su_2021, su_2024_jets, wellons_exploring_2023,byrne_effects_2024}. 

Recent progress in galaxy evolution simulations has allowed for higher hydrodynamic resolution and explicit modeling of different AGN feedback modes, most recently including CRs, along with their dynamical, thermo-chemical, and radiative couplings to gas \citep{hopkins_fire-3_2023}. While some studies have found CRs to be a promising AGN feedback mechanism \citep{Ruszkowski_2017,Su_2019, wellons_exploring_2023}, the details of CR production by AGN and how CRs feedback couples to the host galaxy and surroundings are largely unknown theoretically, and the plausible parameter space remains vast, largely owing to the order-of-magnitude uncertanties in the CR scattering rate (related to sub-AU structure of magnetic fields in halos), which manifests in the CR diffusion/streaming parameters\citep{ruszkowski_cosmic_2023,byrne_effects_2024,ponnada_hooks_2025,hopkins_review_2025}. 

In this study, we explore the orders-of-magnitude variable parameter space of CR feedback from AGN and show that these models reasonably reproduce observed properties of high-mass galaxies (M$_{\rm halo} \approx $ 10$^{13}$ M$_{\odot}$), but crucially differ in their CGM properties, potentially allowing for constraints on these `micro-physical' parameters via emergent observables. We use cosmological zoom-in, cosmic-ray-magneto-hydrodynamic (CR-MHD) galaxy formation simulations from the Feedback In Realistic Environments project (FIRE-3)\footnote{\url{https://fire.northwestern.edu/}} \citep{hopkins_fire-3_2023}, which allows for a more advanced treatment of CR and black hole physics. The relevant physical parameters we vary reflect the uncertainties in CR physics on small, unresolved scales on the scales of CR gyro-radiiin terms of their transport (parameterized by prescriptions for the effective macroscopic scattering rate, which gives rise to diffusion- and streaming-like behaviors) and how efficiently CRs are produced by AGN, i.e., the percentage of AGN accretion energy that is output in the form of CRs.  In Section \ref{sec:methods}, we describe our suite of simulations and analysis methodology before presenting our analysis of the simulations and their properties in Section \ref{sec:results}. Finally, in Section \ref{sec:discussion}, we discuss and conclude our findings.  
\section{Methods} \label{sec:methods}
\subsection{Simulations} \label{subsec:sims}
\begin{deluxetable*}{lcccccc}[ht!]
\tabletypesize{\scriptsize}
\tablewidth{0pt} 
\tablehead{\colhead{Halo} & \colhead{Model} & \colhead{$\epsilon^{\rm BH}_{\rm cr}$ [$10^{-4}$]} & \colhead{log$_{\rm 10}$(M$_{\rm halo}$) [M$_{\odot}$]} & \colhead{log$_{\rm 10}$(M$_{*}$)[M$_{\odot}$]} & \colhead{$E^*_{cr} $ [$10^{58}$ erg]} & \colhead{$E^{BH}_{cr}$ [$10^{58}$ erg]}}
\startdata 
\texttt{h206} & \text{CD} & 3 & 12.7 & 10.9 & 10.74 & 33.22
\\
\hline
\texttt{h206} & \text{VDLoCR} & 1 & 12.7 & 10.9 & 8.78 & 4.94
\\
\hline
\texttt{h206} & \text{VDMidCR} & 10 & 12.7 & 10.8 & 7.03 & 7.12
\\
\hline
\texttt{h206} & \text{VDHiCR} & 30 & 12.7 & 10.4 & 3.21 & 5.15
\\
\hline
\texttt{h113} & \text{CD} & 3 & 12.8 & 11.2 & 19.03 & 63.28
\\
\hline
\texttt{h113} & \text{VDMidCR} & 10 & 12.8 & 11.0 & 12.48 & 9.99
\\
\hline
\texttt{h236} & \text{CD} & 3 & 13.1 & 11.0 & 11.42 & 38.2
\\
\hline
\texttt{h236} & \text{VDMidCR} & 10 & 13.1 & 10.9 & 8.76 & 13.31
\\
\hline
\texttt{h029} & \text{CD} & 3 & 13.0 & 11.4 & 29.5 & 27.95
\\
\hline
\texttt{h029} & \text{VDMidCR} & 10 & 13.0 & 11.1 & 15.9 & 16.28
\enddata 
\caption{Properties of the FIRE-3 simulation suite analyzed in this work. The columns, from left to right, are: name of the halo evolved in FIRE-3, the CR model used to evolve it, energy efficiency of CR injection by the AGN, halo mass, stellar mass, and the total CR energy produced by stars by $z=0$, and the total CR energy produced by the AGN by $z=0$. `VD' models have a variable diffusion coefficient derived from ISM properties whereas `CD' models have a constant power-law diffusion coefficient. The prescription with which stellar and halo mass are defined is described in more detail in Section \ref{subsec:derived_quantities}. The specific prescription used to calculate total CR energy by source is described in detail in Section \ref{subsec:history_methods}. VD runs have less CR production than their CD counterparts, and less AGN feedback relative to the amount of stellar feedback.  \label{tab:sims}}
\end{deluxetable*}
\vspace{-2em}
We analyze a set of FIRE-3 cosmic-ray-magnetohydrodynamic (CR-MHD), cosmological zoom-in simulations, evolved with the \texttt{GIZMO} hydrodynamics solver in meshless-finite-mass mode \citep{hopkins_2015}. All simulations include standard FIRE-3 physics and methods as detailed in \cite{hopkins_fire-3_2023}. We model star formation in self-gravitating, Jeans-unstable gas. Explicit stellar evolution follows feedback arising from radiation pressure, Type I and II supernovae, stellar winds from OB and AGB stars, photoionization, and photo-electric heating. Stellar feedback is coupled to multi-band (extreme UV to far infrared) radiation and gas cooling from 1  - 10$^{10}$ K. This, in conjunction with numerical resolution (M$_{\rm gas}$ $\sim$ 3 $\times\, 10^5$ M$_{\rm \odot}$) naturally gives rise to a multi-phase interstellar and circumgalactic medium.  

All simulations include magnetic fields and anisotropic transport and viscosity, using MHD methods described in \cite{hopkins_&_raives_2016} and \cite{hopkins_2016}. The cosmic ray physics is coupled directly to the magnetohydrodynamics, with the full CR spectrum of MeV-TeV protons and electrons propagated along magnetic field lines according to the fully general CR transport equations, and interacting with the gas via scattering, Lorentz forces, and all relevant loss terms (e.g., radiative, catastrophic, adiabatic, streaming, re-accleration), as described in \citet{hopkins_2022_cr}. CRs from SNe and fast stellar winds are injected with 10\% of the initial SNe/wind kinetic energy and a power-law injection spectrum motivated by diffusive shock acceleration. 

Our simulation suite consists of 4 sets of \textbf{m13} (in reference to their dark matter halo mass M$_{\rm halo} \sim $ 10$^{13}$ M$_{\odot}$) FIRE galaxies. The four m13 halos are \texttt{h206} (some variations of which were explored in \citealt{byrne_effects_2024,ponnada_hooks_2025}), \texttt{h113}, \texttt{h029}, and \texttt{h236}. These simulations all include black holes (BHs) and their explicit associated radiative, mechanical, and cosmic ray feedback modes following \citealt{hopkins_fire-3_2023}; we refer the reader to details therein regarding those feedback channels.

In brief, BH mechanical feedback utilizes a hyper-refined particle spawning method, with particles of $\sim$1000 times higher resolution (lower particle mass) initiated preferentially along the angular momentum axis of the BHs (as perfectly collimated jets) before de-refinement upon mixing into the surrounding gas cells via reaching the local sound speed. All simulations explored in this study utilize initial mechanical wind velocities of 3000 km s$^{-1}$ and a total coupled photon momentum flux (L/c) of 1. 

Each of the four halos was evolved with different CR physics models, varying the energy fraction of AGN accretion that is injected as CRs and the CR transport model. The energy fraction determines the rate of energy injected from the AGN into surrounding gas cells as CRs (with the same injection spectrum as from SNe) whenever mechanical energy is deposited as follows:
$$\dot{E}^{\rm BH}_{\rm cr}  \equiv  \epsilon^{\rm BH}_{\rm cr} \dot{M}_{\rm BH} c^2$$
where $\dot{E}^{\rm BH}_{\rm cr}$ is the CR injection energy rate, $\epsilon^{\rm BH}_{\rm cr}$ is the energy fraction, and $\dot{M}_{\rm BH}$ is the black hole accretion rate \citep{hopkins_fire-3_2023}. 

The two CR transport models we explore are: 1) a temporally and spatially constant power-law scattering rate, which gives rise to an effective diffusion coefficient as a function of CR rigidity (R) of $\kappa_{\rm eff}$ $\sim$ R$^{0.6}$, hereafter referred to as `CD', and 2) a variable diffusion coefficient dependent on the ISM properties, motivated by the ``external driving" model as described in \cite{hopkins_standard_2022} to calibrate scattering rates to reproduce Voyager and AMS-02 observations in Milky Way-mass FIRE-3 simulations, hereafter referred to as `VD'. The VD model is similar in principle to scattering models predicted by ``extrinsic turbulence" models \citep{jokipii_cosmic-ray_1966}, however, with the empirically-motivated addition of a turbulent driving term at gyro-resonant wave-numbers to reproduce the correct spectral shapes in Milky Way Solar Circle-like conditions. This model in principle can produce different scattering and thus CR transport properties from the Milky Way local ISM in different plasma conditions characteristic of the ISM/CGM of massive galaxies explored here. 

We pair the VD model with three CR injection efficiencies: $1 \times 10^{-4}$ (low), $1 \times 10^{-3}$ (medium), and $3 \times 10^{-3}$ (high), hereafter referred to as `VDLoCR', `VDMidCR', and `VDHiCR', respectively. We summarize the simulations analyzed in Table \ref{tab:sims}.

\subsection{Derived Quantities} \label{subsec:derived_quantities}
We calculate several bulk galaxy properties for each simulation, by the same prescriptions as \citet{wellons_exploring_2023} and \citet{ byrne_effects_2024}. Halo mass is defined as the total mass inside the virial radius. Virial radius ($\rm{R_{200}}$) is determined by the radius where the average density equals $200$ times the critical density of the Universe ($\rm{\rho_c = 3 H^2 / 8 \pi G}$). The stellar mass is computed as the total mass of stars within 50 kpc of the galaxy's center, where the galaxy center is defined by the location of the central supermassive black hole (SMBH). For all snapshots, we orient the galaxies face-on using the angular momentum vector of the star particles to define the rotation axis, and transform all cell vector fields accordingly. 

We then measure the 300-Myr averaged star formation rate (SFR) at $z=0$ as the total mass of stars formed in the last 300 Myr within 50 kpc, averaged over that timescale. The effective radius, defined as the two-dimensional radius enclosing half of the stellar mass when the galaxy is viewed face-on, is used to calculate the velocity dispersion of stars, $\sigma$, which is defined as the standard deviation of the $z$-velocity of stars within the effective radius. 

\subsection{Cosmic Ray Injection Histories} \label{subsec:history_methods}
In addition to bulk properties, we calculate the CR injection rates from stellar sources and AGN for each run until $z = 0$. We first calculate the time-series SFR as the total mass of star particles formed within 50 kpc of the galactic center, corrected for mean mass loss rates to obtain the at-formation stellar mass \citep{hopkins_fire-3_2023}. From there, we approximate the stellar CR injection rate as $\dot{E}^*_{\rm cr} \approx 0.1\times10^{51}\text{ erg }(\dot{M}_* / 100 M_\odot )$, where $0.1\times10^{51}\text{ erg }$ is the amount of CR energy released from a single supernova event, which occurs approximately once per $100\, M_\odot$ of the stars formed. To obtain the cumulative CR energy injected into the galaxy by stellar feedback, we integrate the injection rate over time. We use the AGN accretion rate to calculate the CR injection rate from the AGN, where $\dot{E}^{\rm BH}_{\rm cr}  \equiv  \epsilon^{\rm BH}_{\rm cr} \dot{M}_{\rm BH} c^2\approx 1.8\times10^{59} \text{ erg } (\dot{M}_{\rm BH} / 10^8 M_\odot)(\epsilon^{\rm BH}_{\rm cr} / 10^{-3})$. We cumulatively integrate the time-series injection rates over time for each source to find the cumulative CR energy injected into the galaxy by each source.  

\subsection{Radial Profiles} \label{subsec:profile_methods}
Finally, we analyze various radial profiles for each simulation snapshot. 
The CR energy density profiles at different redshifts are computed as the total CR energy contained in shells of $\sim$1.7 kpc thickness at a given radius away from the galaxy's center, divided by the volume of each shell. Following a similar procedure, we also generate gas number density and effective CR diffusion coefficient ($\kappa_{\rm eff}$) profiles for different redshifts. To calculate $\kappa_{\rm eff}$, we first average the diffusivity of the $\sim$0.1-100 GeV proton spectrum by taking the energy-weighted mean of the diffusivities of the spectral bins for each given cell. Then, we use the aforementioned volume-weighting to obtain a radial profile of $\kappa_{\rm eff}$. 

\section{Results} \label{sec:results}
\subsection{Bulk Properties}
\label{subsec:bulk_properties}

\begin{figure} 
    \centering
    \includegraphics[width=.37\textwidth, trim=5em 0em 5em 4em]{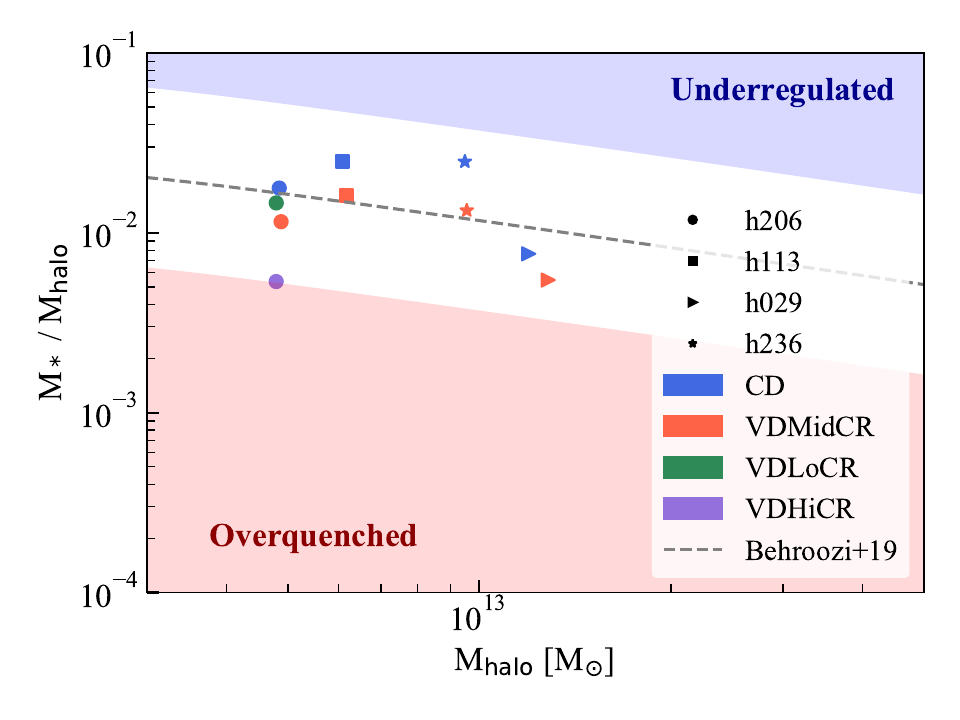}
    \includegraphics[width=.37\textwidth, trim=5em 0em 5em 2em]{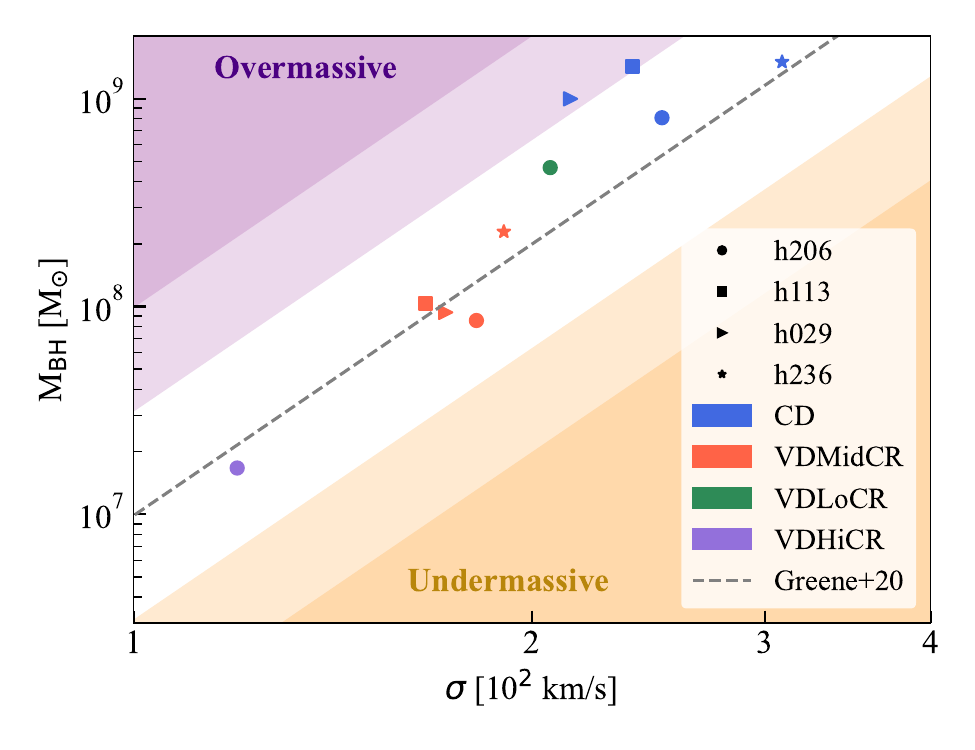}
    \includegraphics[width=.37\textwidth, trim=5em 2em 5em 2em]{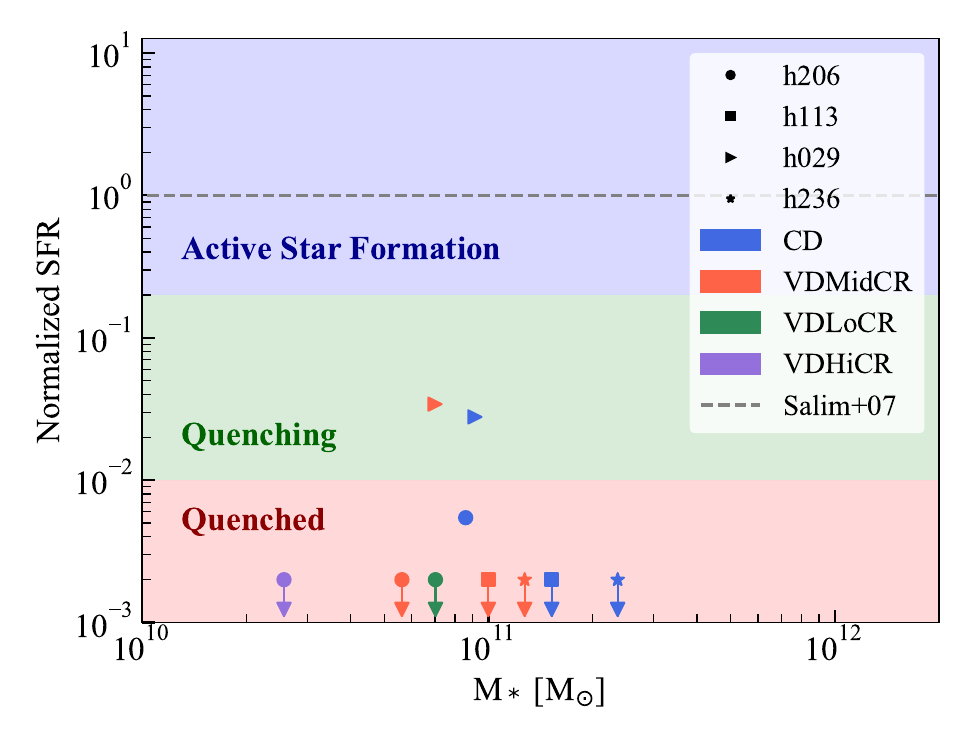}
    \caption{Scaling relations for all \texttt{h206}, \texttt{h113}, \texttt{h029}, and \texttt{h236} runs at $z$ = 0, indicated by squares, circles, triangles, and stars, respectively. Blue, green, red, and purple markers indicate runs with the CD, VDLoCR, VDMidCR, and VDHiCR model, respectively. Irrespective of the details of CR physics, all models produce well-regulated, quenched galaxies with reasonable SMBH masses, in agreement with the scaling relations. \textbf{\textit{Top:}} The stellar mass - halo mass relation from \citet{Behroozi_2019} (dashed line). The blue and red regions describes galaxies with under-regulated and overquenched stellar populations  respectively. \textbf{\textit{Center:}} The M$_{\rm BH}$-$\sigma$ relation from \citet{Greene_2020} (dashed line). The purple and green regions describes galaxies with overly-massive and undermassive AGN respectively. \textbf{\textit{Bottom:}} The 300-Myr averaged star formation rate, normalized to the median value of star formation observed in star-forming galaxies at $z$ = 0 from \citet{Salim_2007}. The blue, green, and red regions describe actively star forming, quenching, and quenched galaxies. }
\label{fig:bulk}
\end{figure}

The bulk properties of the galaxies are plotted in Figure \ref{fig:bulk}, and show that the models produce galaxies that are in agreement with known empirical scaling relations. 
\\ \indent
The top panel shows the stellar mass - halo mass (SMHM) relation, tracking the amount of star formation regulation. The galaxies are compared to the median observational SMHM found by \citet{Behroozi_2019} of central galaxies (both star-forming and quenched), following \citet{wellons_exploring_2023}, with galaxies within 0.5 dex of the line being considered as properly regulated. All models explored here produce galaxies that fall within the range of being well-regulated. As might be expected, higher cosmic-ray injection efficiency (stronger feedback) leads to increased regulation of the stellar mass. VD models produce more regulated galaxies in comparison to the CD model, including the VDLoCR run of \texttt{h206}, despite having a lower CR injection efficiency than CD. We also note that $M_{\text{halo}}$ has little variation across models for each halo, producing a qualitatively distinct vertical band for each halo in the panel.  
\\ \indent
In the middle panel of Figure \ref{fig:bulk}, we show the $\text{M}_{\text{BH}}$-$\sigma$ relation at $z=0$, where $\sigma$ refers to the velocity dispersion of stars, as described in Section \ref{subsec:derived_quantities}. The models are compared with the observational relation of \citet{Greene_2020}, with the model SMBH considered to be properly massive if the mass is within 0.5 dex of the observational constraint. All models we explore also follow this relation. Models with higher cosmic-ray injection efficiencies are seen producing galaxies with lower AGN mass and velocity dispersion, suggesting that the stronger the CR feedback, the more self-regulation of AGN growth, corresponding with the same trend as seen in the SMHM relation. However, the CD model produces galaxies with the highest black hole mass despite having higher CR efficiency than VDLoCR, suggesting that variable CR transport is more efficient at regulating BH growth for a given energy injection, potentially via stronger confinement of the CRs in the inner CGM. The $\text{M}_{\text{BH}}$-$\sigma$ relation also appears to be less sensitive to variations in initial halo conditions, producing a cluster for each specific model across different halos. 
\\ \indent
The bottom panel shows the 300-Myr averaged SFR of each model variation and halo at $z=0$, normalized to the median star formation expected from a star forming main sequence galaxy (including AGN/SF composite galaxies) at $z=0$ as found by \citet{Salim_2007}, against the stellar mass to measure star formation suppression. All runs are within 1-2 dex below the expected SFR for actively star forming galaxies, indicating that the galaxies are mildly quenched. VD models appear to have marginally higher $z=0$ SFRs but lower stellar masses than their CD counterparts for the same halo. Thus, we find that all CR parameterizations explored yield reasonably quenched galaxies with appropriate stellar and black hole masses at this halo mass scale.

\subsection{Cosmic Ray Injection Histories \& Implications for Quenching}
\begin{figure*}[ht]
    \centering
    \includegraphics[width=.9\textwidth]{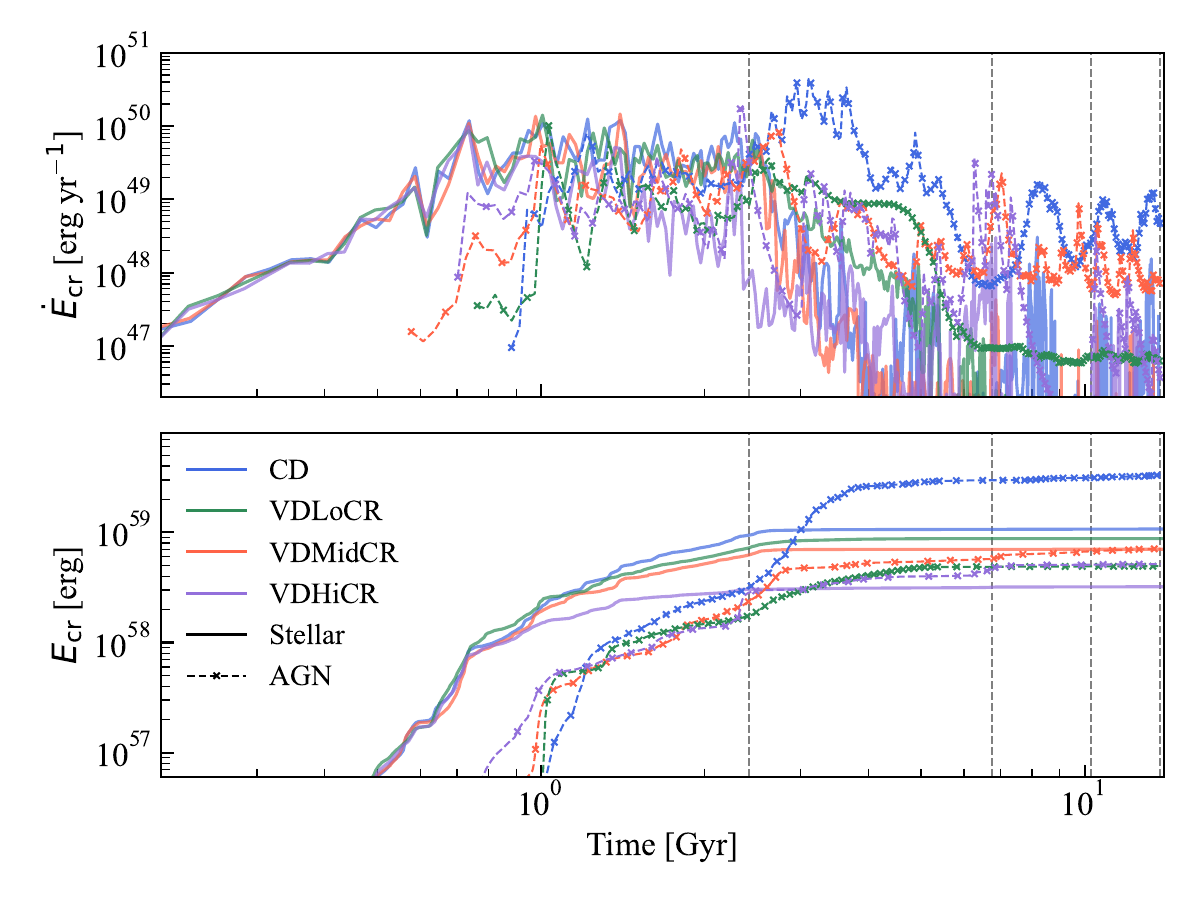}
    \caption{
    Cosmic ray injection rate (top) and cumulative injected energy (bottom) by source of the \texttt{h206} runs, from 0 to 13.8 Gyr. Stellar feedback is plotted in solid; AGN feedback is plotted in dashed lines with `x' markers. Vertical dashed lines correspond to snapshots showcased below in Figures \ref{fig:CR_density} - \ref{fig:kappa}. The injection rate is binned at $\sim$30 Myr intervals for both stellar and AGN injection. Stellar injection rate is relatively steady-state after $\sim$6 Gyr with respect to star-forming galaxies. All runs exhibit a moving average in AGN injection rate. VDMidCR has the earliest AGN-projected CR injection and has both the highest injection rate and cumulative injection at 13.8 Gyr compared to the other VD runs. Unlike other runs, VDLoCR's AGN injection rate stabilizes beyond $\sim$6 Gyr.}
    \label{fig:injection_by_source}
    
\end{figure*} 
Similarly to \cite{byrne_effects_2024}, we approximate the total energy of CRs produced from stars over the age of the universe as $E^*_{\rm cr} \approx 0.1\times10^{51}\text{ erg }(M_{\rm sf} / 100 M_\odot )$, where $0.1\times10^{51}\text{ erg }$ is the amount of CR energy released from a single supernova event, which occurs approximately once per $100\, M_\odot$ of the stars formed, and $M_{\rm sf}$ is the cumulative at-formation stellar mass (see Section \ref{subsec:profile_methods} for more detailed methodology). Similarly, we approximate the total CR energy produced by the AGN as $E^{\rm BH}_{\rm cr} = \epsilon^{\rm BH}_{\rm cr} M_{\rm BH, acc} c^2 \approx 1.8\times10^{59} \text{ erg } (M_{\rm BH, acc} / 10^8 M_\odot)(\epsilon^{\rm BH}_{\rm cr} / 10^{-3})$ where $M_{\rm BH, acc}$ is the cumulative mass accreted onto the AGN. Table \ref{tab:sims} summarizes the cumulative CR energy deposited from each source by $z=0$ for all runs in our simulation suite.

We find that while VD runs have at most $\sim$1.6 times more CR energy from AGN feedback compared to stellar feedback, the CD runs have up to $\sim$3.3 times the CR energy from AGN vs. stellar sources. The CD runs produce more CR energy from both stars and the AGN compared to VDMidCR across all halos, despite having a lower injection efficiency. 

This is because the VD transport models control BH growth more effectively than CD. Moreover, this indicates that the primary way in which the increased CR feedback efficiency regulates galaxy growth is not via aggregate injection of CRs, but rather due to the \textit{timing} of injection, which is inherently chaotic and sensitive to the `responsiveness' of the BH feedback model, as described in \citet{wellons_exploring_2023}.

\begin{figure*}[ht!]
    \centering
    \includegraphics[width=.95\textwidth]{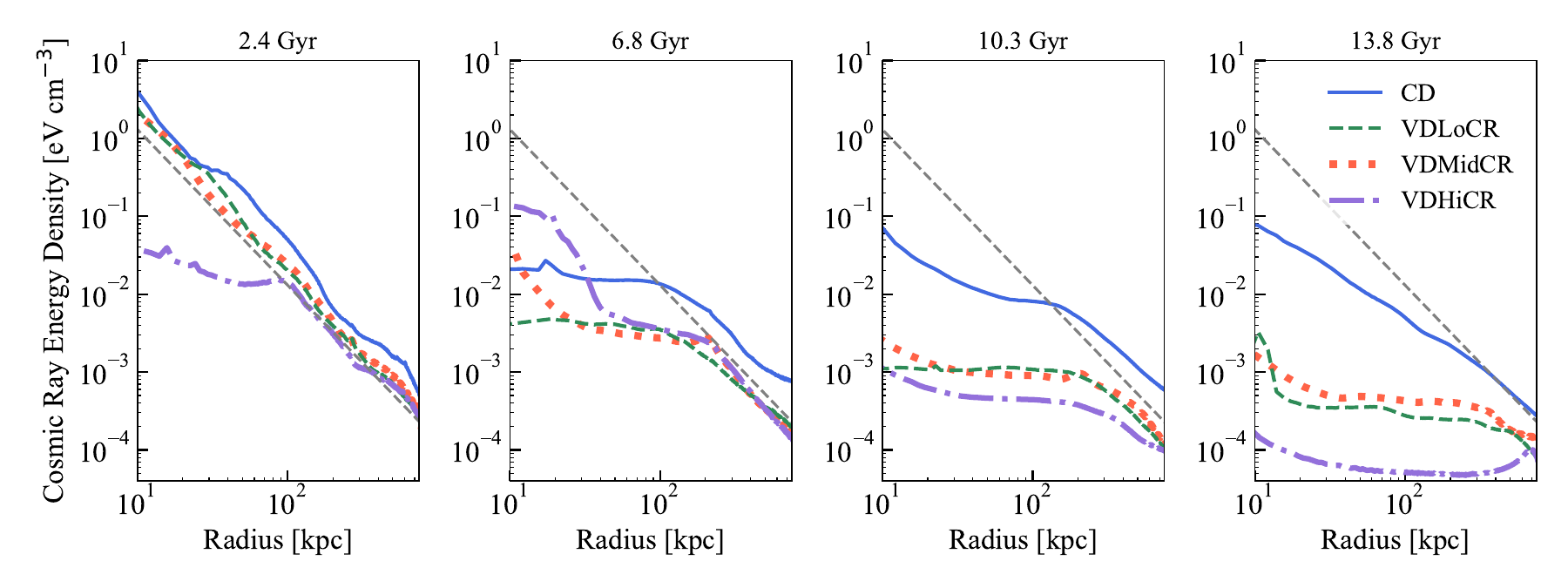}
    \caption{Radial profiles of the volume-weighted mean e$_{\rm cr}$ in the halos of the \texttt{h206} runs, binned in $\sim$1.7 kpc spherical annuli. Snapshots at $\tau = 2.4, 6.8, 10.3, 13.8$ Gyr are shown in the left to right panels respectively. The gray dashed line is the $1/r^2$ profile expected from steady state CR injection and $\kappa_{\rm eff} \propto r$ \citep{Butsky_2023}. VD models have increasingly flatter profiles over time, with a nonlinear relationship between the normalization of the e$_{\rm cr}$ profiles and CR injection efficiencies ($\epsilon^{\rm BH}_{\rm cr}$).}
    \label{fig:CR_density}
\end{figure*} 

Interestingly, we also find that while $E^{*}_{\rm cr}$ decreases linearly with CR injection efficiency in \texttt{h206}'s VD runs, $E^{\rm BH}_{\rm cr}$ peaks with VDMidCR, highlighting the same non-linear behavior of increasing CR feedback efficiency as seen in the CR energy density profiles in the CGM at $z=0$ (Figure \ref{fig:CR_density}). This potentially indicates a threshold-like behavior from a largely feedback-regulated to a mostly fueling-regulated regime.

We calculate the CR injection rates from stellar and AGN feedback for the \texttt{h206} runs and present them in Figure \ref{fig:injection_by_source}. We find increasing variability in AGN injection rates for VD models with increasing injection efficiency, indicating higher responsiveness of the feedback model. The CD model produces AGN injection rates systematically higher than those of VDLoCR after $\sim$5 Gyr despite having similar profiles before then. Note again, the CD model has a factor of 3 higher CR injection efficiency than that of VDLoCR and a factor of $\sim$6.7 larger total E$^{BH}_{\rm cr}$ and despite this, the VDLoCR model exhibits stronger regulation on the scaling relations (Figure \ref{fig:bulk}). This indicates that the CR transport variation alone can play a significant role in the responsiveness of a given AGN feedback model.

The VDMidCR and VDHiCR runs have slightly earlier initial AGN injection times and exhibit higher variation in AGN injection rates at late times, but still on average lower than that of the CD model. As expected, stellar CR injection rates are relatively low compared to star-forming galaxies after quenching past $\sim$6 Gyr, with brief bursts typically sub-dominant to the BH contribution. This indicates all models primarily regulate via AGN feedback, and more so that they are regulated via its responsiveness, here particularly that of the CR feedback model \citep{wellons_exploring_2023}.

We note here also that the self-regulation our simulated galaxies exhibit is non-trivial. In Fig. \ref{fig:bulk}, we showed that for increasing CR feedback efficiency, \texttt{h206} runs \textit{moved down} the $z=0$ SMHM relation while moving \textit{along} the M$_{\rm BH}$-$\sigma$ relation. In the feedback-regulated BH growth picture \citep{di_matteo_energy_2005,sijacki_unified_2007,sijacki_growing_2009}, accretion onto the BH gets shut down due to `quasar-mode' feedback, which is more ejective \& explosive in nature. This is evinced by the similar E$^{\rm BH}_{cr}$ injected by each VD model, and their subsequent M$_{\rm BH}$ which corresponds inversely to $\epsilon^{\rm BH}_{\rm cr}$.

Subsequently, evolution enters a `fueling-regulated' mode (commonly referred to or implemented as `radio-mode' feedback; \citealt{weinberger_supermassive_2018}) where M$_{\rm BH}$ changes slowly and star-formation is modulated by prevention of gas accretion. This `maintenance-mode' allows for M$_{\ast}$ to further change, modifying the central potential and thus $\sigma$, moving points down the SMHM relation and along the M$_{\rm BH}$-$\sigma$ relation. We have verified this qualitative behavior by examining the M$_{\rm BH}$-$\sigma$ relation for each individual snapshot.

Importantly, in our simulations here, we have not inserted `by hand' any qualitative change in the feedback prescription for varied Eddington ratios as in large-volume cosmological simulations -- the halos' collective response to feedback naturally arises from the coupling of feedback to gas, which varies in its responsiveness, and the cosmological assembly of the galaxies. This highlights how multi-channel AGN feedback, varied solely in CR efficiency, can naturally produce \textit{both} qualitative feedback behaviors necessary to reproduce bulk massive galaxy observables.

\subsection{CGM Properties}
\label{subsec:radial_profiles_results}

\begin{figure*}[t]
    \centering
    \includegraphics[width=1\textwidth]{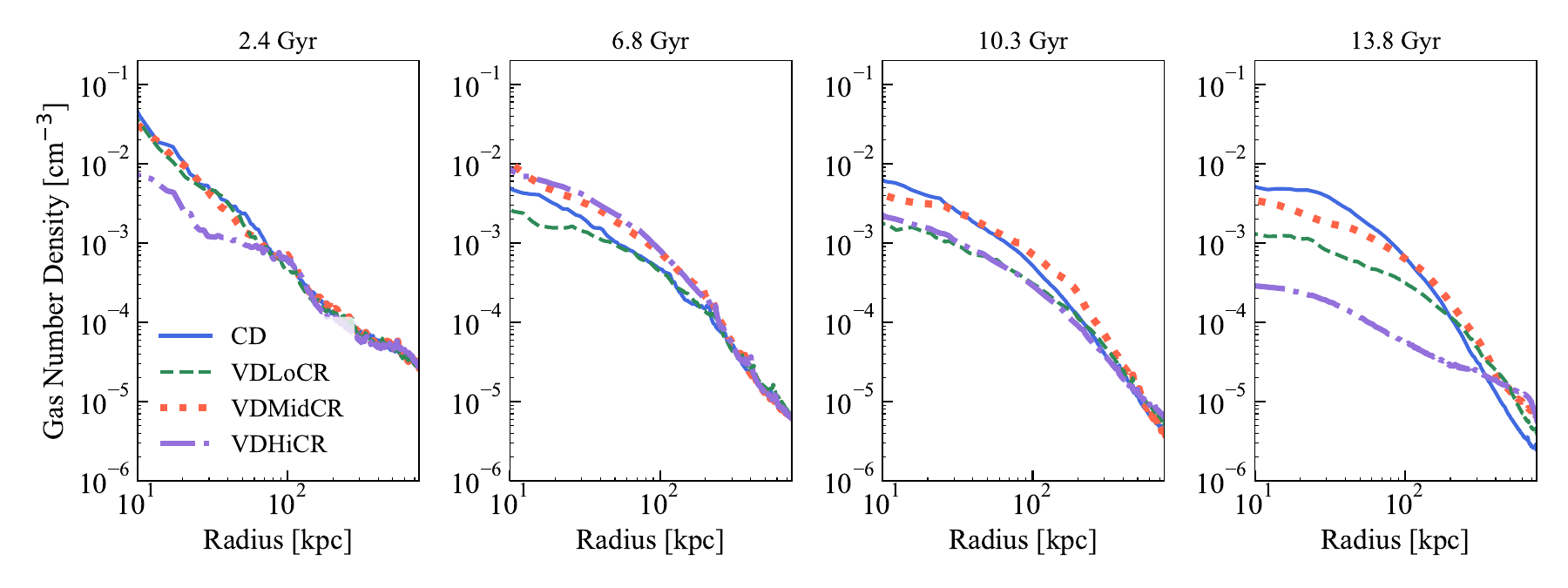}
    \caption{Radial profiles of the volume-weighted mean n$_{\rm gas}$ for the \texttt{h206} runs, from 10 kpc to 750 kpc binned in $\sim$1.7 kpc spherical annuli. Snapshots at $\tau = 2.4, 6.8, 10.3, 13.8$ Gyr are shown in the left to right panels respectively. Similar trends as the CR density profile suggest dynamical coupling of CRs to gas.}
    \label{fig:n_density}
\end{figure*}

\begin{figure*}
    \centering
    \includegraphics[width=1\textwidth]{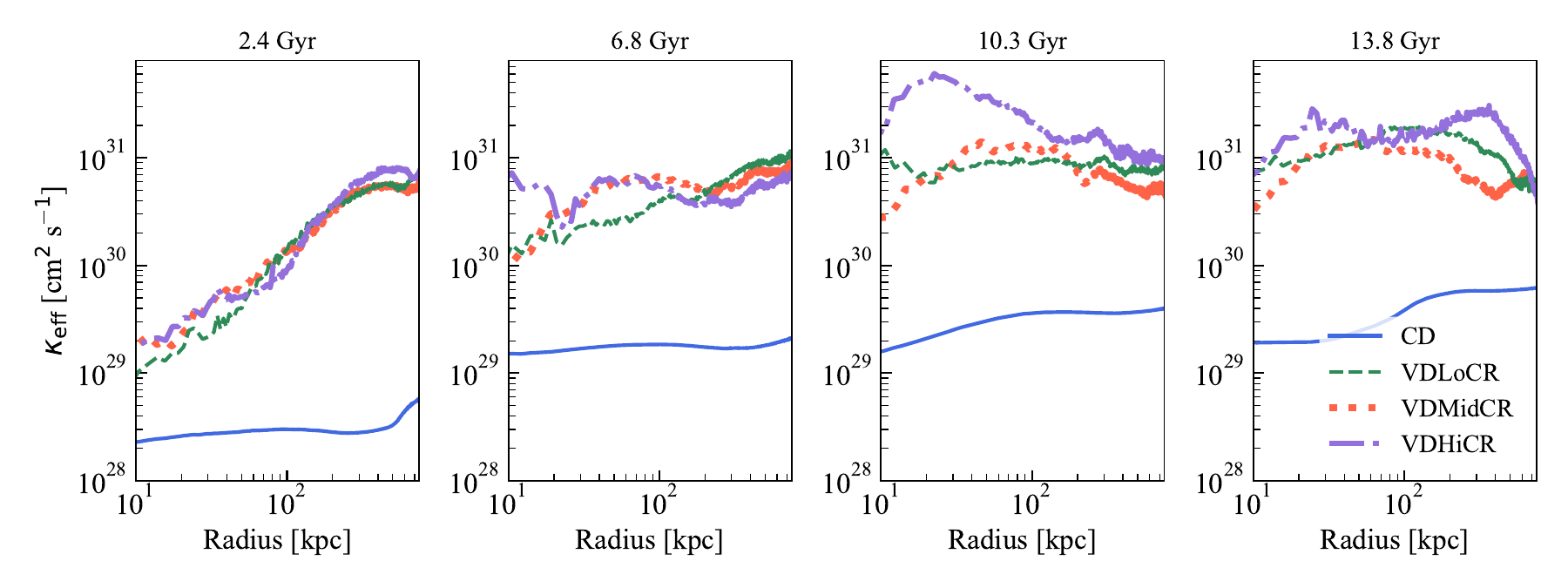}
    \caption{Radial profiles of the volume-weighted mean cosmic ray diffusivity $\kappa_{\text{eff}}$ in the halos of the \texttt{h206} runs, from 10 kpc to 750 kpc binned in $\sim$1.7 kpc annuli. Snapshots at $\tau = 2.4, 6.8, 10.3, 13.8$ Gyr are shown in the left to right panels respectively.
    The presences of troughs suggests the possibility of CR winds. }
    \label{fig:kappa}
\end{figure*} 
  
In this subsection, we present radial profiles of various physical quantities in the CGM from our simulations. We find that all halos produce similar qualitative profile trends for a given physics variation, so we focus on \texttt{h206} hereafter. The radial profiles of CR energy density are shown in Figure \ref{fig:CR_density}. Different models exhibit different radial profiles of CR energy density e$_{\rm cr}$, or equivalently the CR pressure since P$_{\rm cr} \approx $$\frac{1}{3}$e$_{\rm cr}$ for $\gamma_{\rm cr}$ = 4/3. In particular, CD and VD models evolve differently with time, with the CD model presenting a power-law-like profile at late times and VD models exhibiting qualitatively flatter radial profiles, particularly at late times. For the VD models of varied injection efficiency, we find that the CR energy density profiles in the CGM at $z=0$ exhibit a nonlinear ordering, where the VDMidCR model has the highest e$_{\rm cr}$ amongst all VD models at radii $\gtrsim$15 kpc. The VDHiCR model produces the largest variation in its e$_{\rm cr}$ profile over time, having the lowest e$_{\rm cr}$ in the inner halo at the earliest time ($\tau = 2.4\,  \rm Gyr$) followed by the highest inner halo e$_{\rm cr}$ at $\tau$ = 6.8 Gyr among the VD models and then decreasing to the lowest energy density at $z=0$, with its CGM e$_{\rm cr}$ nearly an order of magnitude lower at a given radius than that of VDMidCR and VDLoCR models at $z=0$.

The VDHiCR model exhibits a similar trend over time in its radial profile of gas number density (Figure \ref{fig:n_density}). At $z=0$, the gas density radial profiles share the same ordering with CR injection efficiency as CR energy density does; in general, CGM gas density profiles appear to flatten over time irrespective of the model variation, particularly within $r \lesssim 300 \, \rm kpc$ (out to the approximate virial radii of these halos). The similarities between the CGM gas density and CR energy density owe to the dynamical coupling of CRs to the gas, as steep gradients in e$_{\rm cr}$ at the edge of flattened ``shelves'' in the profiles aid in outflows. Since the CRs can diffuse, and the CR pressure is time-dependent, the shelves in e$_{\rm cr}$ can move ahead of the gas \citep{hopkins_CR_winds_Mpc_2021,ponnada_2025_time_dependent}, and so e$_{\rm cr}$ as a function of radius (or gas density) does not appear to trace an exact, tightly-coupled adiabat \citep{hopkins_2022_cr}. Despite this caveat, and even though these halos are not CR pressure dominated overall, there appears to be some degree of spatial correspondence between the gradual flattening of e$_{\rm cr}$ with time shown in Fig. \ref{fig:CR_density} and the qualitative dilution and flattening of the gas density profiles with time in Figure \ref{fig:n_density}, particularly in the VDHiCR case which has the largest modulation in AGN-injected e$_{\rm cr}$ over time.

In Figure \ref{fig:kappa}, we show the radial profiles of the volume-weighted mean CR \textit{effective} diffusion coefficient (which is itself energy-averaged over all CR energy bins for each gas cell; see Section \ref{sec:methods}), $\kappa_{\text{eff}}$, at different redshifts. Here, $\kappa_{\text{eff}}$ contains both `diffusion-like' and `streaming-like' terms which emerge from the evolved scattering rate as a function of rigidity, $\nu_{\rm cr} (\rm R),$ but \textit{not} the advective transport of CRs. The CD model, as expected, has a largely constant effective diffusivity, with slight changes representing small shifts in the CR energy spectrum from the canonical $\sim \rm GeV$ peak at a given radius. The ``troughs'' present in the VD profiles correspond to areas of lower diffusivity surrounded by high diffusivity that could trap CRs, thus creating $e_{\rm cr}$ overdensities. Indeed, we see such troughs in $\kappa_{\rm eff}$ correspond to overdensities in $e_{\rm cr}$ in Figures \ref{fig:CR_density} and \ref{fig:kappa} e.g. in VDHiCR at $\sim$15 and $\sim$200 at $\tau = 6.8$ Gyr; in VDMidCR at $\sim$200 kpc at $\tau = 10.3$ Gyr and $\sim$400 kpc at $\tau = 13.8$ Gyr; and in VDLoCR at $\sim$15 kpc at $\tau = 10.3$ Gyr. Sharper troughs correspond to stronger local CR pressure gradients and overdensities. 

We note here that at the earliest times just after the period of the most rapid BH growth (for instance at $\tau = 2.4\, \rm Gyr$), all of the models' CGM properties in Figures \ref{fig:CR_density}-\ref{fig:kappa} are more similar than they are different, which is particularly true of the VD runs, though with some differences in the inner e$_{\rm CR}$ profiles set by the earliest 'feedback-regulated' phase. Much of the differences in the CGM properties at $z \sim0$ are instead set by the behavior of the models at later times during the `maintenance-mode' phase of growth. 

\section{Discussion and Conclusions} \label{sec:discussion}

In this work, we have explored plausible models for AGN feedback with varied CR injection efficiencies from the black hole, and two different CR transport models. All parameterizations of CR physics explored here produce reasonable galaxies with suppressed star formation rates, suggesting that bulk properties of the galaxies are relatively insensitive to CR parameterizations. However, the radial profiles of CR energy density and gas properties reveal differences in model predictions, which may reveal an avenue to constrain how AGN feedback regulates galaxies, and particularly, the role of CRs in quenching.

The CR energy density profiles are flatter than expected from a constant injection of CRs and steady-state halo CR pressure \citep{Butsky_2023}, which owes to the time-dependent nature of the injection. \citet{ponnada_2025_time_dependent} show that time-dependent injection of CRs (decreasing $\dot{E}_{\rm cr}$ with time) undergoing a mix of diffusive- and streaming/advection-like transport leads to flattening of P$_{\rm cr}$(r) relative to time-steady expectations. 

Indeed, this behavior is shown in \citet{ponnada_2025_time_dependent} for the same CD run of \texttt{m13h206} we analyze here and agrees well with time-dependent analytic expectations. The CGM n$_{\rm gas}$ profiles also appear to flatten similarly, which may indicate that the CRs are playing a role in driving outflows via their dynamical coupling to the gas. 

In Figure \ref{fig:injection_by_source}, we detailed the fine-grained CR injection rates from AGN and star-formation for each \texttt{h206} run. These largely show a decreasing $\langle\dot{E}_{\rm cr}\rangle$ with time, where $\langle \rangle$ denotes averaging over some effective transport timescale out to CGM radii. However, there is significant burstiness on short timescales, which depending on the balance of diffusive- vs. streaming/advection-like transport will be ``smeared out" to varying degrees.

Our current understanding of CR transport physics, which is critical for modulating CGM phase structure, is limited by the lack of direct observables that track CRs and low-density CGM. Moreover, $\epsilon^{\rm BH}_{\rm cr}$ is a highly uncertain quantity, and may also vary in terms of hadronic/leptonic composition \citep{lin_evolution_2023}. 

FIRE-2 simulations evolved with single-bin treatments of CRs with different physically-motivated sub-grid scattering prescriptions have produced varying CGM properties \citep{ponnada2024synchrotronsignaturescosmicray, hopkins_2021_cr_model_effects}, as have FIRE-3 galaxies evolved with AGN feedback and varied multi-bin CR transport physics (including some explored herein), specifically differing in morphological evolution of the far-infrared-radio correlation \citep{ponnada_hooks_2025}, which may be independently constraining for CR effects in the ISM and inner CGM when compared to spatially resolved observations \citep{Murphy_2006}.

Recently, \citep{hopkins_cosmic_2025} proposed inverse Compton emission from CR leptons as a method by which to estimate the CR lepton pressure, which is largely tracing the AGN-produced CRs for these simulations at late times (at least within $\sim 1-2\,R_{\rm vir}$ for the $\kappa_{\rm eff}$ modeled here). The distinctions between the n$_{\rm gas}$ and P$_{\rm cr}$ profiles for different $\epsilon^{\rm BH}_{\rm cr}$ (for a fixed CR transport model) appear promising towards constraining $\epsilon^{\rm BH}_{\rm cr}$, but a more statistical sample of simulations and corresponding synthetic observations would be needed to compare against existing constraints, which are often stacked measurements (e.g., from eROSITA \citealt{zhang_hot_2024,zhang_hot_2024-1,zhang_hot_2025}). 

There is also the complementary probe of thermal pressure in the CGM via the thermal Sunyaev-Zel'dovich effect, which when combined with X-ray constraints may be particularly constraining how CRs contribute to galaxy quenching \citep{ponnada_strong_2025}. Quenched massive galaxies appear to be \textit{brighter} in X-rays than their star-forming counterparts with the latest detections \citep{zhang_hot_2025}, while simulations invoking AGN feedback without CRs tend to over-predict the detected halo-integrated tSZ signal by over an order-of-magnitude \citep{das_thermal_2025,ponnada_strong_2025}. 

So, by producing synthetic synchrotron \citep{ponnada_multibin_synchrotron_emission}, IR-O/UV \citep{byrne_effects_2024,wijers_ne_2024,qutob_observational_2024,ponnada_hooks_2025}, and X-ray predictions \citep{chadayammuri_testing_2022,lu_constraining_2025} for feedback parameterized with varied CR injection efficiencies and transport parameterizations, our results suggest we may be able to constrain these physics via observational comparisons of spectrally-resolved CR-MHD simulations with explicit evolution of multi-channel AGN feedback. While the AGN feedback schemes explored herein have injected the multi-channel feedback energy and momentum on the accretion-kernel scale ($\sim 1-10 \,\rm pc$ from the central source), recent developments using idealized setups have suggested that alternative modes of CR injection, e.g. at jet termination shocks \citep{su_modeling_2025}, may qualitatively change the quenching and maintenance behavior of a fixed set of feedback parameters in massive galaxies. Our results motivate further exploring models as presented herein and in \citet{su_modeling_2025} with explicit feedback in a large number of cosmological runs to place detailed constraints on AGN feedback. 

Finally, we note that structures we see arising in P$_{\rm cr}$(r) due to time-dependence as well as variation in $\kappa_{\rm eff}$ at large radii may potentially connect to formation of Odd Radio Circles (ORCs) observed in radio surveys around massive galaxies \citep{norris_odd_2021}.  
Since we have only explored one plausible model for spatially variable CR scattering, we do not perform a comprehensive comparison here, and leave it to future work.\newline

Support for CG, SP, and PFH was provided by a Simons Investigator Award. SW received support from NASA grant 80NSSC24K0838.


\vspace{5mm}
\facilities{TACC}


\software{astropy \citep{2013A&A...558A..33A}}



\newpage
\bibliography{AGN_CRs}{}

\begin{thebibliography}{}
\expandafter\ifx\csname natexlab\endcsname\relax\def\natexlab#1{#1}\fi
\providecommand{\url}[1]{\href{#1}{#1}}
\providecommand{\dodoi}[1]{doi:~\href{http://doi.org/#1}{\nolinkurl{#1}}}
\providecommand{\doeprint}[1]{\href{http://ascl.net/#1}{\nolinkurl{http://ascl.net/#1}}}
\providecommand{\doarXiv}[1]{\href{https://arxiv.org/abs/#1}{\nolinkurl{https://arxiv.org/abs/#1}}}

\bibitem[{{Astropy Collaboration} {et~al.}(2013){Astropy Collaboration}, {Robitaille}, {Tollerud}, {Greenfield}, {Droettboom}, {Bray}, {Aldcroft}, {Davis}, {Ginsburg}, {Price-Whelan}, {Kerzendorf}, {Conley}, {Crighton}, {Barbary}, {Muna}, {Ferguson}, {Grollier}, {Parikh}, {Nair}, {Unther}, {Deil}, {Woillez}, {Conseil}, {Kramer}, {Turner}, {Singer}, {Fox}, {Weaver}, {Zabalza}, {Edwards}, {Azalee Bostroem}, {Burke}, {Casey}, {Crawford}, {Dencheva}, {Ely}, {Jenness}, {Labrie}, {Lim}, {Pierfederici}, {Pontzen}, {Ptak}, {Refsdal}, {Servillat}, \& {Streicher}}]{2013A&A...558A..33A}
{Astropy Collaboration}, {Robitaille}, T.~P., {Tollerud}, E.~J., {et~al.} 2013, \aap, 558, A33, \dodoi{10.1051/0004-6361/201322068}

\bibitem[{Behroozi {et~al.}(2019)Behroozi, Wechsler, Hearin, \& Conroy}]{Behroozi_2019}
Behroozi, P., Wechsler, R.~H., Hearin, A.~P., \& Conroy, C. 2019, Monthly Notices of the Royal Astronomical Society, 488, 3143, \dodoi{10.1093/mnras/stz1182}

\bibitem[{Benson(2010)}]{Benson_2010}
Benson, A.~J. 2010, Galaxy formation theory,  Elsevier {BV}, \dodoi{10.1016/j.physrep.2010.06.001}

\bibitem[{Bustard \& Zweibel(2021)}]{bustard_cosmic-ray_2021}
Bustard, C., \& Zweibel, E.~G. 2021, The Astrophysical Journal, 913, 106, \dodoi{10.3847/1538-4357/abf64c}

\bibitem[{Butsky {et~al.}(2023)Butsky, Nakum, Ponnada, Hummels, Ji, \& Hopkins}]{Butsky_2023}
Butsky, I.~S., Nakum, S., Ponnada, S.~B., {et~al.} 2023, Monthly Notices of the Royal Astronomical Society, 521, 2477, \dodoi{10.1093/mnras/stad671}

\bibitem[{Butsky \& Quinn(2018)}]{butsky_role_2018}
Butsky, I.~S., \& Quinn, T.~R. 2018, The Astrophysical Journal, 868, 108, \dodoi{10.3847/1538-4357/aaeac2}

\bibitem[{Byrne {et~al.}(2024)Byrne, Faucher-Giguère, Wellons, Hopkins, Anglés-Alcázar, Sultan, Wijers, Moreno, \& Ponnada}]{byrne_effects_2024}
Byrne, L., Faucher-Giguère, C.-A., Wellons, S., {et~al.} 2024, The Astrophysical Journal, 973, 149, \dodoi{10.3847/1538-4357/ad67ca}

\bibitem[{Chadayammuri {et~al.}(2022)Chadayammuri, Bogdan, Oppenheimer, Kraft, Forman, \& Jones}]{chadayammuri_testing_2022}
Chadayammuri, U., Bogdan, A., Oppenheimer, B.~D., {et~al.} 2022, The Astrophysical Journal Letters, 936, L15, \dodoi{10.3847/2041-8213/ac8936}

\bibitem[{Cole {et~al.}(2002)Cole, Lacey, Baugh, \& Frenk}]{Cole_2002}
Cole, S., Lacey, C.~G., Baugh, C.~M., \& Frenk, C.~S. 2002, Monthly Notices of the Royal Astronomical Society, 319, 168, \dodoi{10.1046/j.1365-8711.2000.03879.x}

\bibitem[{{Croton} \& {Farrar}(2008)}]{Croton2008}
{Croton}, D.~J., \& {Farrar}, G.~R. 2008, \mnras, 386, 2285, \dodoi{10.1111/j.1365-2966.2008.13204.x}

\bibitem[{{Croton} {et~al.}(2006){Croton}, {Springel}, {White}, {De Lucia}, {Frenk}, {Gao}, {Jenkins}, {Kauffmann}, {Navarro}, \& {Yoshida}}]{Croton2006}
{Croton}, D.~J., {Springel}, V., {White}, S. D.~M., {et~al.} 2006, \mnras, 365, 11, \dodoi{10.1111/j.1365-2966.2005.09675.x}

\bibitem[{Das {et~al.}(2025)Das, Truong, Chiang, \& Mathur}]{das_thermal_2025}
Das, S., Truong, N., Chiang, Y.-K., \& Mathur, S. 2025, The Astrophysical Journal, 991, 205, \dodoi{10.3847/1538-4357/adfdd6}

\bibitem[{Davé {et~al.}(2019)Davé, Anglés-Alcázar, Narayanan, Li, Rafieferantsoa, \& Appleby}]{Dav_2019}
Davé, R., Anglés-Alcázar, D., Narayanan, D., {et~al.} 2019, Monthly Notices of the Royal Astronomical Society, 486, 2827–2849, \dodoi{10.1093/mnras/stz937}

\bibitem[{Di~Matteo {et~al.}(2005)Di~Matteo, Springel, \& Hernquist}]{di_matteo_energy_2005}
Di~Matteo, T., Springel, V., \& Hernquist, L. 2005, Nature, 433, 604, \dodoi{10.1038/nature03335}

\bibitem[{Fabian(2012)}]{fabian_observational_2012}
Fabian, A.~C. 2012, Annual Review of Astronomy and Astrophysics, 50, 455, \dodoi{10.1146/annurev-astro-081811-125521}

\bibitem[{Greene {et~al.}(2020)Greene, Strader, \& Ho}]{Greene_2020}
Greene, J.~E., Strader, J., \& Ho, L.~C. 2020, Annual Review of Astronomy and Astrophysics, 58, 257, \dodoi{10.1146/annurev-astro-032620-021835}

\bibitem[{Harrison \& Almeida(2024)}]{harrison_observational_2024}
Harrison, C.~M., \& Almeida, C.~R. 2024, Observational {Tests} of {Active} {Galactic} {Nuclei} {Feedback}: {An} {Overview} of {Approaches} and {Interpretation},  arXiv.
\newblock \url{http://arxiv.org/abs/2404.08050}

\bibitem[{Harrison {et~al.}(2018)Harrison, Costa, Tadhunter, Flütsch, Kakkad, Perna, \& Vietri}]{Harrison_2018}
Harrison, C.~M., Costa, T., Tadhunter, C.~N., {et~al.} 2018, Nature Astronomy, 2, 198, \dodoi{10.1038/s41550-018-0403-6}

\bibitem[{Heckman \& Best(2014)}]{heckman_coevolution_2014}
Heckman, T.~M., \& Best, P.~N. 2014, Annual Review of Astronomy and Astrophysics, 52, 589, \dodoi{10.1146/annurev-astro-081913-035722}

\bibitem[{Hopkins(2015)}]{hopkins_2015}
Hopkins, P.~F. 2015, Monthly Notices of the Royal Astronomical Society, 450, 53, \dodoi{10.1093/mnras/stv195}

\bibitem[{Hopkins(2025)}]{hopkins_review_2025}
---. 2025, Cosmic {Rays} on {Galaxy} {Scales}: {Progress} and {Pitfalls} for {CR}-{MHD} {Dynamical} {Models},  arXiv, \dodoi{10.48550/arXiv.2509.07104}

\bibitem[{Hopkins {et~al.}(2022{\natexlab{a}})Hopkins, Butsky, Panopoulou, Ji, Quataert, Faucher-Giguère, \& Kereš}]{hopkins_2022_cr}
Hopkins, P.~F., Butsky, I.~S., Panopoulou, G.~V., {et~al.} 2022{\natexlab{a}}, Monthly Notices of the Royal Astronomical Society, 516, 3470, \dodoi{10.1093/mnras/stac1791}

\bibitem[{Hopkins {et~al.}(2021)Hopkins, Chan, Ji, Hummels, Kereš, Quataert, \& Faucher-Giguère}]{hopkins_CR_winds_Mpc_2021}
Hopkins, P.~F., Chan, T.~K., Ji, S., {et~al.} 2021, Monthly Notices of the Royal Astronomical Society, 501, 3640, \dodoi{10.1093/mnras/staa3690}

\bibitem[{Hopkins {et~al.}(2020{\natexlab{a}})Hopkins, Chan, Squire, Quataert, Ji, Kereš, \& Faucher-Giguère}]{hopkins_2021_cr_model_effects}
Hopkins, P.~F., Chan, T.~K., Squire, J., {et~al.} 2020{\natexlab{a}}, Monthly Notices of the Royal Astronomical Society, 501, 3663, \dodoi{10.1093/mnras/staa3692}

\bibitem[{Hopkins {et~al.}(2025)Hopkins, Quataert, Ponnada, \& Silich}]{hopkins_cosmic_2025}
Hopkins, P.~F., Quataert, E., Ponnada, S.~B., \& Silich, E. 2025, The Open Journal of Astrophysics, 8, 78, \dodoi{10.33232/001c.141293}

\bibitem[{Hopkins \& Raives(2015)}]{hopkins_&_raives_2016}
Hopkins, P.~F., \& Raives, M.~J. 2015, Monthly Notices of the Royal Astronomical Society, 455, 51, \dodoi{10.1093/mnras/stv2180}

\bibitem[{Hopkins {et~al.}(2022{\natexlab{b}})Hopkins, Squire, Butsky, \& Ji}]{hopkins_standard_2022}
Hopkins, P.~F., Squire, J., Butsky, I.~S., \& Ji, S. 2022{\natexlab{b}}, Monthly Notices of the Royal Astronomical Society, \dodoi{10.1093/mnras/stac2909}

\bibitem[{Hopkins {et~al.}(2016)Hopkins, Torrey, Faucher-Giguère, Quataert, \& Murray}]{hopkins_2016}
Hopkins, P.~F., Torrey, P., Faucher-Giguère, C.-A., Quataert, E., \& Murray, N. 2016, Monthly Notices of the Royal Astronomical Society, 458, 816, \dodoi{10.1093/mnras/stw289}

\bibitem[{{Hopkins} {et~al.}(2004){Hopkins}, {Strauss}, {Hall}, {Richards}, {Cooper}, {Schneider}, {Vanden Berk}, {Jester}, {Brinkmann}, \& {Szokoly}}]{Hopkins2004}
{Hopkins}, P.~F., {Strauss}, M.~A., {Hall}, P.~B., {et~al.} 2004, \aj, 128, 1112, \dodoi{10.1086/423291}

\bibitem[{Hopkins {et~al.}(2018)Hopkins, Wetzel, Kere{\v{s}}, Faucher-Gigu{\`{e}}re, Quataert, Boylan-Kolchin, Murray, Hayward, Garrison-Kimmel, Hummels, Feldmann, Torrey, Ma, Angl{\'{e}}s-Alc{\'{a}}zar, Su, Orr, Schmitz, Escala, Sanderson, Grudi{\'{c}}, Hafen, Kim, Fitts, Bullock, Wheeler, Chan, Elbert, \& Narayanan}]{Hopkins_2018}
Hopkins, P.~F., Wetzel, A., Kere{\v{s}}, D., {et~al.} 2018, Monthly Notices of the Royal Astronomical Society, 480, 800, \dodoi{10.1093/mnras/sty1690}

\bibitem[{Hopkins {et~al.}(2020{\natexlab{b}})Hopkins, Chan, Garrison-Kimmel, Ji, Su, Hummels, Kereš, Quataert, \& Faucher-Giguère}]{Hopkins2020}
Hopkins, P.~F., Chan, T.~K., Garrison-Kimmel, S., {et~al.} 2020{\natexlab{b}}, Monthly Notices of the Royal Astronomical Society, 3465, \dodoi{10.1093/mnras/stz3321}

\bibitem[{Hopkins {et~al.}(2023)Hopkins, Wetzel, Wheeler, Sanderson, Grudić, Sameie, Boylan-Kolchin, Orr, Ma, Faucher-Giguère, Kereš, Quataert, Su, Moreno, Feldmann, Bullock, Loebman, Anglés-Alcázar, Stern, Necib, Choban, \& Hayward}]{hopkins_fire-3_2023}
Hopkins, P.~F., Wetzel, A., Wheeler, C., {et~al.} 2023, Monthly Notices of the Royal Astronomical Society, 519, 3154, \dodoi{10.1093/mnras/stac3489}

\bibitem[{Jokipii(1966)}]{jokipii_cosmic-ray_1966}
Jokipii, J.~R. 1966, The Astrophysical Journal, 146, 480, \dodoi{10.1086/148912}

\bibitem[{Kereš {et~al.}(2009)Kereš, Katz, Davé, Fardal, \& Weinberg}]{Keres_2009}
Kereš, D., Katz, N., Davé, R., Fardal, M., \& Weinberg, D.~H. 2009, Monthly Notices of the Royal Astronomical Society, 396, 2332–2344, \dodoi{10.1111/j.1365-2966.2009.14924.x}

\bibitem[{Lin {et~al.}(2023)Lin, Yang, \& Owen}]{lin_evolution_2023}
Lin, Y.-H., Yang, H.-Y.~K., \& Owen, E.~R. 2023, Monthly Notices of the Royal Astronomical Society, 520, 963, \dodoi{10.1093/mnras/stad185}

\bibitem[{Longair(2008)}]{Longair:2008gba}
Longair, M.~S. 2008, {Galaxy Formation}, Astronomy and Astrophysics Library (Heidelberg, Germany: Springer), \dodoi{10.1007/978-3-540-73478-9}

\bibitem[{Lu {et~al.}(2025)Lu, Kereš, Hopkins, Ponnada, Faucher-Giguére, \& Hummels}]{lu_constraining_2025}
Lu, Y.~S., Kereš, D., Hopkins, P.~F., {et~al.} 2025, Constraining cosmic ray transport models using circumgalactic medium properties and observables,  arXiv, \dodoi{10.48550/arXiv.2505.13597}

\bibitem[{Murphy {et~al.}(2006)Murphy, Braun, Helou, Armus, Kenney, Gordon, Bendo, Dale, Walter, Oosterloo, Kennicutt, Calzetti, Cannon, Draine, Engelbracht, Hollenbach, Jarrett, Kewley, Leitherer, Li, Meyer, Regan, Rieke, Rieke, Roussel, Sheth, Smith, \& Thornley}]{Murphy_2006}
Murphy, E.~J., Braun, R., Helou, G., {et~al.} 2006, The Astrophysical Journal, 638, 157–175, \dodoi{10.1086/498636}

\bibitem[{Norris {et~al.}(2021)Norris, Crawford, \& Macgregor}]{norris_odd_2021}
Norris, R.~P., Crawford, E., \& Macgregor, P. 2021, Galaxies, 9, 83, \dodoi{10.3390/galaxies9040083}

\bibitem[{Ponnada(2025)}]{ponnada_2025_time_dependent}
Ponnada, S.~B. 2025, Time-{Dependent} {Cosmic} {Ray} {Halos} from {Bursty} {Star} {Formation} and {Active} {Galactic} {Nuclei}: {Semi}-{Analytic} {Formalism} and {Galaxy} {Formation} {Implications},  arXiv, \dodoi{10.48550/arXiv.2509.02697}

\bibitem[{Ponnada {et~al.}(2025{\natexlab{a}})Ponnada, Hopkins, Lu, Silich, Butsky, \& Keres}]{ponnada_strong_2025}
Ponnada, S.~B., Hopkins, P.~F., Lu, Y.~S., {et~al.} 2025{\natexlab{a}}, Strong {Evidence} for {Cosmic} {Ray}-{Supported} \${\textbackslash}sim\${L}\${\textasciicircum}\{{\textbackslash}ast\}\$ {Galaxy} {Halos} via {X}-ray {\textbackslash}\& {tSZ} {Constraints},  arXiv, \dodoi{10.48550/arXiv.2510.13959}

\bibitem[{Ponnada {et~al.}(2023)Ponnada, Panopoulou, Butsky, Hopkins, Skalidis, Hummels, Quataert, Kereš, Faucher-Giguère, \& Su}]{ponnada_multibin_synchrotron_emission}
Ponnada, S.~B., Panopoulou, G.~V., Butsky, I.~S., {et~al.} 2023, Monthly Notices of the Royal Astronomical Society, 527, 11707–11718, \dodoi{10.1093/mnras/stad3978}

\bibitem[{Ponnada {et~al.}(2024)Ponnada, Butsky, Skalidis, Hopkins, Panopoulou, Hummels, Kereš, Quataert, Faucher-Giguère, \& Su}]{ponnada2024synchrotronsignaturescosmicray}
Ponnada, S.~B., Butsky, I.~S., Skalidis, R., {et~al.} 2024, Synchrotron Signatures of Cosmic Ray Transport Physics in Galaxies.
\newblock \doarXiv{2309.16752}

\bibitem[{Ponnada {et~al.}(2025{\natexlab{b}})Ponnada, Cochrane, Hopkins, Butsky, Wellons, Sanchez, Hummels, Lu, Kereš, \& Hayward}]{ponnada_hooks_2025}
Ponnada, S.~B., Cochrane, R.~K., Hopkins, P.~F., {et~al.} 2025{\natexlab{b}}, The Astrophysical Journal, 980, 135, \dodoi{10.3847/1538-4357/ada280}

\bibitem[{Qutob {et~al.}(2024)Qutob, Emami, Su, Smith, Hernquist, Triani, Hummels, Fielding, Hopkins, Somerville, Ballantyne, Vogelsberger, Tremblay, Steiner, Finkbeiner, Narayan, Park, Grindlay, Natarajan, Hayward, Kereš, Ponnada, Belli, Davies, Maheson, Bugiani, \& Li}]{qutob_observational_2024}
Qutob, N., Emami, R., Su, K.-Y., {et~al.} 2024, The Astrophysical Journal, 977, 72, \dodoi{10.3847/1538-4357/ad8658}

\bibitem[{Ruszkowski \& Pfrommer(2023)}]{ruszkowski_cosmic_2023}
Ruszkowski, M., \& Pfrommer, C. 2023, Astronomy and Astrophysics Review, 31, 4, \dodoi{10.1007/s00159-023-00149-2}

\bibitem[{Ruszkowski {et~al.}(2017)Ruszkowski, Yang, \& Reynolds}]{Ruszkowski_2017}
Ruszkowski, M., Yang, H.-Y.~K., \& Reynolds, C.~S. 2017, The Astrophysical Journal, 844, 13, \dodoi{10.3847/1538-4357/aa79f8}

\bibitem[{Salim {et~al.}(2007)Salim, Rich, Charlot, Brinchmann, Johnson, Schiminovich, Seibert, Mallery, Heckman, Forster, Friedman, Martin, Morrissey, Neff, Small, Wyder, Bianchi, Donas, Lee, Madore, Milliard, Szalay, Welsh, \& Yi}]{Salim_2007}
Salim, S., Rich, R.~M., Charlot, S., {et~al.} 2007, The Astrophysical Journal Supplement Series, 173, 267, \dodoi{10.1086/519218}

\bibitem[{Schaye {et~al.}(2014)Schaye, Crain, Bower, Furlong, Schaller, Theuns, Dalla~Vecchia, Frenk, McCarthy, Helly, Jenkins, Rosas-Guevara, White, Baes, Booth, Camps, Navarro, Qu, Rahmati, Sawala, Thomas, \& Trayford}]{Schaye_2014}
Schaye, J., Crain, R.~A., Bower, R.~G., {et~al.} 2014, Monthly Notices of the Royal Astronomical Society, 446, 521–554, \dodoi{10.1093/mnras/stu2058}

\bibitem[{Sijacki {et~al.}(2007)Sijacki, Springel, Di~Matteo, \& Hernquist}]{sijacki_unified_2007}
Sijacki, D., Springel, V., Di~Matteo, T., \& Hernquist, L. 2007, Monthly Notices of the Royal Astronomical Society, 380, 877, \dodoi{10.1111/j.1365-2966.2007.12153.x}

\bibitem[{Sijacki {et~al.}(2009)Sijacki, Springel, \& Haehnelt}]{sijacki_growing_2009}
Sijacki, D., Springel, V., \& Haehnelt, M.~G. 2009, Monthly Notices of the Royal Astronomical Society, 400, 100, \dodoi{10.1111/j.1365-2966.2009.15452.x}

\bibitem[{Sijacki {et~al.}(2015)Sijacki, Vogelsberger, Genel, Springel, Torrey, Snyder, Nelson, \& Hernquist}]{Sijacki_2015}
Sijacki, D., Vogelsberger, M., Genel, S., {et~al.} 2015, Monthly Notices of the Royal Astronomical Society, 452, 575–596, \dodoi{10.1093/mnras/stv1340}

\bibitem[{{Silk} \& {Rees}(1998)}]{SilkandRees1998}
{Silk}, J., \& {Rees}, M.~J. 1998, \aap, 331, L1, \dodoi{10.48550/arXiv.astro-ph/9801013}

\bibitem[{Somerville \& Dav\'{e}(2015)}]{Somerville_&_Dave_2015}
Somerville, R.~S., \& Dav\'{e}, R. 2015, Annual Review of Astronomy and Astrophysics, 53, 51, \dodoi{10.1146/annurev-astro-082812-140951}

\bibitem[{Springel \& Hernquist(2003)}]{Springel_2003}
Springel, V., \& Hernquist, L. 2003, Monthly Notices of the Royal Astronomical Society, 339, 312, \dodoi{10.1046/j.1365-8711.2003.06207.x}

\bibitem[{Su {et~al.}(2025)Su, Bryan, Hopkins, Natarajan, Ponnada, Emami, \& Lu}]{su_modeling_2025}
Su, K.-Y., Bryan, G.~L., Hopkins, P.~F., {et~al.} 2025, Modeling {Cosmic} {Rays} at {AGN} {Jet}-{Driven} {Shock} {Fronts},  arXiv, \dodoi{10.48550/arXiv.2502.00927}

\bibitem[{Su {et~al.}(2019{\natexlab{a}})Su, Hopkins, Hayward, Ma, Faucher-Giguère, Kereš, Orr, Chan, \& Robles}]{su_failure_2019}
Su, K.-Y., Hopkins, P.~F., Hayward, C.~C., {et~al.} 2019{\natexlab{a}}, Monthly Notices of the Royal Astronomical Society, 487, 4393, \dodoi{10.1093/mnras/stz1494}

\bibitem[{Su {et~al.}(2019{\natexlab{b}})Su, Hopkins, Hayward, Faucher-Gigu{\`{e}}re, Kere{\v{s}}, Ma, Orr, Chan, \& Robles}]{Su_2019}
---. 2019{\natexlab{b}}, Cosmic rays or turbulence can suppress cooling flows (where thermal heating or momentum injection fail),  Oxford University Press ({OUP}), \dodoi{10.1093/mnras/stz3011}

\bibitem[{Su {et~al.}(2021)Su, Hopkins, Bryan, Somerville, Hayward, Angl{\'{e} }s-Alc{\'{a}}zar, Faucher-Gigu{\`{e}}re, Wellons, Stern, Terrazas, Chan, Orr, Hummels, Feldmann, \& Kere{\v{s}}}]{Su_2021}
Su, K.-Y., Hopkins, P.~F., Bryan, G.~L., {et~al.} 2021, Monthly Notices of the Royal Astronomical Society, 507, 175, \dodoi{10.1093/mnras/stab2021}

\bibitem[{{Su} {et~al.}(2024){Su}, {Bryan}, {Hayward}, {Somerville}, {Hopkins}, {Emami}, {Faucher-Gigu{\`e}re}, {Quataert}, {Ponnada}, {Fielding}, \& {Kere{\v{s}}}}]{su_2024_jets}
{Su}, K.-Y., {Bryan}, G.~L., {Hayward}, C.~C., {et~al.} 2024, \mnras, 532, 2724, \dodoi{10.1093/mnras/stae1629}

\bibitem[{Weinberger {et~al.}(2018)Weinberger, Springel, Pakmor, Nelson, Genel, Pillepich, Vogelsberger, Marinacci, Naiman, Torrey, \& Hernquist}]{weinberger_supermassive_2018}
Weinberger, R., Springel, V., Pakmor, R., {et~al.} 2018, Monthly Notices of the Royal Astronomical Society, 479, 4056, \dodoi{10.1093/mnras/sty1733}

\bibitem[{Wellons {et~al.}(2023)Wellons, Faucher-Giguère, Hopkins, Quataert, Anglés-Alcázar, Feldmann, Hayward, Kereš, Su, \& Wetzel}]{wellons_exploring_2023}
Wellons, S., Faucher-Giguère, C.-A., Hopkins, P.~F., {et~al.} 2023, Monthly Notices of the Royal Astronomical Society, 520, 5394, \dodoi{10.1093/mnras/stad511}

\bibitem[{Wijers {et~al.}(2024)Wijers, Faucher-Giguère, Stern, Byrne, \& Sultan}]{wijers_ne_2024}
Wijers, N.~A., Faucher-Giguère, C.-A., Stern, J., Byrne, L., \& Sultan, I. 2024, The Astrophysical Journal, 973, 99, \dodoi{10.3847/1538-4357/ad63a0}

\bibitem[{Zhang {et~al.}(2024{\natexlab{a}})Zhang, Comparat, Ponti, Merloni, Nandra, Haberl, Locatelli, Zhang, Sanders, Zheng, Liu, Popesso, Liu, Truong, Pillepich, Predehl, Salvato, Shreeram, Yeung, \& Ni}]{zhang_hot_2024}
Zhang, Y., Comparat, J., Ponti, G., {et~al.} 2024{\natexlab{a}}, Astronomy \& Astrophysics, 690, A267, \dodoi{10.1051/0004-6361/202449412}

\bibitem[{Zhang {et~al.}(2024{\natexlab{b}})Zhang, Comparat, Ponti, Merloni, Nandra, Haberl, Truong, Pillepich, Locatelli, Zhang, Sanders, Zheng, Liu, Popesso, Liu, Predehl, Salvato, Shreeram, \& Yeung}]{zhang_hot_2024-1}
---. 2024{\natexlab{b}}, Astronomy \& Astrophysics, 690, A268, \dodoi{10.1051/0004-6361/202449413}

\bibitem[{Zhang {et~al.}(2025)Zhang, Comparat, Ponti, Merloni, Nandra, Haberl, Truong, Pillepich, Popesso, Locatelli, Zhang, Sanders, Zheng, Liu, Liu, Predehl, Salvato, Bruggen, Shreeram, \& Yeung}]{zhang_hot_2025}
---. 2025, Astronomy \& Astrophysics, 693, A197, \dodoi{10.1051/0004-6361/202452273}

\end{thebibliography}
\bibliographystyle{aasjournal}



\end{document}